\documentclass[prd,aps,nofootinbib]{revtex4}
\usepackage{amssymb}
\usepackage{graphicx}
\usepackage{amsmath}
\usepackage{epsfig}

\begin{document}

\title{\textbf{Approach to equilibrium for a class of random quantum models of infinite range}}

\date{\today}

\author{Walter F. Wreszinski$^{(a)}$}\email[]{wreszins@gmail.com}

\affiliation{(a) Universidade de Sao Paulo, Sao Paulo, Brazil}

\vspace{5.0cm}

\begin{abstract}
We consider random generalizations of a quantum model of infinite range introduced by Emch and Radin. The generalizations allow a neat extension from the class $l_1$ of absolutely summable lattice potentials to the optimal class 
$l_2$ of square summable potentials first considered by Khanin and Sinai and generalised by van Enter and van Hemmen. The approach to equilibrium in the case of a Gaussian distribution is proved to be faster than for a Bernoulli distribution for both short-range and long-range lattice potentials. While exponential decay to equilibrium is excluded in the nonrandom $l_1$ case, it is proved to occur for both short and long range potentials for Gaussian distributions, and for potentials of class $l_2$ in the Bernoulli case. Open problems are discussed.
\end{abstract}

\maketitle
\newpage
\section{Introduction and Summary}
\vspace{1.0cm}
Equilibrium properties of general random quantum systems have been the subject of recent investigations [1]. Mean-field models, e.g., quantum analogues of the Sherrington-Kirkpatrick (SK) spin-glass have also been studied, and the thermodynamic limit proved for the SK model in a transverse external field [2]. In this paper we consider the dynamics of a special class of random quantum models of infinite range, which may be viewed as a version with random couplings of the model introduced by Emch [3] and generalized by Radin [4].

The most compelling reason for studying random quantum systems - in particular, quantum spin glasses - is that only for a quantum system does a physically satisfactory definition of the dynamics of states and observables exist. Kinetic theory approaches to the SK spin glass model, for instance, rely on Glauber dynamics [5], which is an imposed dynamics: indeed, the Ising model is, of course - both mathematically and physically - to be regarded as an anisotropic limit of suitable quantum systems, whose dynamics is naturally defined. The occurrence of ageing and other non-equilibrium phenomena were also derived under the assumption of an imposed dynamics, such as the Langevin dynamics for the spherical and mean spherical model
with competing interactions treated in ([6],[7]).

The Emch-Radin model seems to be largely ignored in the recent literature. Indeed, in an influential paper on the mathematical theory of non-equilibrium statistical mechanics [24], Jaksic and Pillet remark that, to their knowledge, the dynamical aspects of quantum spin systems have been little studied, the only well-understood case being the one-dimensional XY model [25]. An additional point lending special interest to the model of ([3],[4]) is that it has had a  qualitatively successful application to the description of real experiments, which display a non-Markovian approach to equilibrium ([27],[28]), as well as the fact that several results are not restricted to dimension one ([3],[4]) . From a purely conceptual point of view, the quantum extension of the Kac ring model in [26] should also be mentioned.
 
We now comment on the organization of the paper.

Section II reviews the nonrandom model ([3],[4]). We show that the infinite volume limit 
$\langle\sigma^{x}_{i_0}\rangle(t)$of the one-site transverse spin exists if the lattice potential is square-summable 
(class $l_2$). For the case in which the lattice potential is absolutely summable (class $l_1$), for no form of the interaction the approach to equilibrium is of exponential type: this is a well-known result in function theory, mentioned for completeness in proposition 2.1, because reference is made to it later.

Section III introduces the random models, which furnish a natural interpretation of the result of section II for square integrable lattice potentials because they are, in contrast to the nonrandom case, thermodynamically stable. This was first observed for classical models by Khanin and Sinai [8] and extended, with an independent proof, to quantum systems by van Enter and van Hemmen [9]. For definiteness, we restrict ourselves throughout the paper, to two opposite extreme examples of probability distributions, viz. Bernoulli and Gaussian, and show that the Bernoulli rate of approach to equilibrium coincides with the nonrandom case. We also prove that the mean transverse magnetization converges along a subsequence, almost everywhere in the probability space, to its average, a nonrandom function $f(t)$.

In section IV we show that the approach to equilibrium is always faster than Bernoulli ,and exponential, for the Gaussian distribution, both for short range and long range potentials of classes $l_1$ and $l_2$, while for Bernoulli it is exponential only for potentials of class $l_2$.

Section V is reserved to a conclusion and open problems.

\section{The nonrandom model}  

Following Radin's review [4] of the work of Emch [3], consider the experiment of [9] (see also [10] ): a $CaF_2$ crystal is placed in a magnetic field thus determining the z-direction, and allowed to reach thermal equilibrium. A rf pulse is then applied which turns the net nuclear magnetization to the $x$ direction. The magnetization in the $x$ direction is then measured as a function of time.

As in ([3],[4]), we assume an interaction of the form
$$
H_{V}=1/2\sum_{j,k}\epsilon(\vert j-k \vert)\sigma^{z}_{j}\sigma^{z}_{k} - B\sum_{j}\sigma^{z}_{j}
\eqno{(1)}
$$
where $j\in V$ and $k\in V$ in (1), where $V$ is a finite region, $V\in\mathbf{Z}^{\nu}$. $H_{V}$ is defined on the Hilbert space 
$$
\bigotimes_{i\in V} \mathbf{C}^{2}_{i}
\eqno{(2)}
$$
The state representing the system after the application of the rf pulse will be assumed to be the product state
$$
\rho=\rho(0)=\bigotimes_{j\in V}\phi_{j}
\eqno{(3a)}
$$
where
$$
\phi_{j}(\cdot)=tr_{j}(\cdot \exp(-\gamma\sigma^{x}_{j}))/tr _{j}(\exp(-\gamma\sigma^{x}_{j})
\eqno{(3b)}
$$
where $tr$ is the trace on $\mathbf{C}^{2}_{j}$. Other choices of the state are possible [4], but we shall adopt (3) for definiteness. Let
$$
S^{x}=1/N(V)\sum_{j\in V}\sigma^{x}_{j}
\eqno{(4)}
$$
be the mean transverse magnetization, with $N(V)$ denoting the number of sites in $V$. The real number $\gamma$ in (3b) may be chosen as in [3] such as to maximize the microscopic entropy subject to the constraint $tr S^{x}\rho(0)$ equal to a constant, i.e., a given value of the mean transverse magnetization. Since the state (3a) is a product state, $\gamma$ is independent of $V$, and has the same value if $S^{x}$ is replaced by $\sigma^{x}_{i_0}$, for any $i_0\in V$. Let
$$
\rho^{V}_{t}\equiv U_{V}^{t} \rho(0)U_{V}^{-t}
\eqno{(5a)}
$$
and
$$
\langle\sigma^{x}_{i_0}\rangle_{V}(t)\equiv \rho^{V}_{t}(\sigma^{x}_{i_0})
\eqno{(5b)}
$$
where $U_{V}(t)=\exp(itH_{V})$. We may write, by (5),
$$
\langle\sigma^{x}_{i_0}\rangle_{V}(t)=\rho(0)(U_{V}(-t)\sigma^{x}_{i_0}U_{V}(t))
\eqno{(6)}
$$
It is natural to define
$$
\langle\sigma^{x}_{i_0}\rangle(t)=\lim_{V\to\infty}\langle\sigma^{x}_{i_0}\rangle_{V}(t)
\eqno{(7)}
$$
provided the limit on the r.h.s. of (7) exists, as the expectation value of the local transverse spin in the (time-dependent) nonequilibrium state $\rho_{\infty}^{t}$ of the infinite system. A weak-star limit [11] of the sequence of states 
$\rho_{V}^{t}(\cdot)= tr(\cdot U_{V}^{t}\rho(0)U_{V}^{-t})$ exists, by compactness, on the usual quasilocal algebra $\cal{A}$  of observables. The expectation value of $\sigma^{x}_{i_0}$ in the equilibrium state associated to (1) is zero by the symmetry of rotation by $\pi$ around the $z$ axis. We may now pose the question whether the limit
$$
f(t)\equiv \lim_{N(\Lambda)\to\infty}\rho_{\infty}^{t}(1/N(\Lambda)\sum_{i_0\in\Lambda}\sigma^{x}_{i_0})
\eqno{(8)}
$$
where $\Lambda$ denotes a finite subset of $\mathbf{Z}^{\nu}$, which is interpreted as the mean transverse magnetization, exists. The property of approach to equilibrium is expressed  by 
$$
\lim_{t\to\infty}f(t)=0
\eqno{(9)}
$$  

Of particular interest is the rate of approach to equilibrium. In the present nonrandom case, the limit at the r.h.s. of (8) equals $f(t)=\rho_{\infty}^{t}(\sigma^{x}_{i_0})$, for any $i_0$, by translation invariance of $\rho_{\infty}^{t}$. This is not so for random systems, in which case additional arguments are necessary to show the convergence of the r.h.s. of (8) for almost all configurations of couplings (section III).

At this point we simplify the analysis as in [3] by restricting ourselves to the one-dimensional case. Generalization of some results to any dimension $\nu$ is straightforward, as remarked in [3], because they rely on the general Proposition 1 of ref. [4]. The Hamiltonian becomes
$$
H_n=H_{V=[-n,n]}=1/2\sum_{j,k=-n}^{n}\epsilon(\vert j-k \vert)\sigma^{z}_{j}\sigma^{z}_{k}+B\sum_{j=-n}^{n}\sigma^{z}_{j}
\eqno{(10)}
$$
where the function $\epsilon(\cdot)$ is assumed to satisfy:
$$
\epsilon(0)=0
\eqno{(11a)}
$$
$$
\epsilon(n) \mbox{ is monotonically decreasing in } n
\eqno{(11b)}
$$
and either:
$$
\sum_{n=1}^{\infty} \vert \epsilon(n) \vert < \infty \mbox{ i.e. } \epsilon\in l_1
\eqno{(12a)}
$$
or
$$
\sum_{n=1}^{\infty} (\vert \epsilon(n) \vert)^{2} < \infty  \mbox{ i.e. } \epsilon\in l_2
\eqno{(12b)}
$$

The terms containing the interaction $\epsilon(j)$ in the operator $U_{n}^{-t}\sigma^{x}_{i_0}U_{n}^{t}$ ($n=V$) which appear on the r.h.s. of (6) are of the form (see [3], pg.1200):
$$
\lbrack \cos(\epsilon(j)t)-i\sigma^{z}_{i_0}\sigma^{z}_{i_0-j}\sin(\epsilon(j)t)\rbrack\times\\
\lbrack \sigma^{x}_{i_0}\cos(2\epsilon(j)t)+\sigma^{y}_{i_0}\sigma^{z}_{i_0+j}\sin(2\epsilon(j)t)\rbrack\times\\
\lbrack \cos(\epsilon(j)t)+i\sigma^{z}_{i_0}\sigma^{z}_{i_0-j}\sin(\epsilon(j)t)\rbrack
\eqno{(14)}
$$
By (3), and emplying a product basis of eigenstates
$$
\Psi_{\alpha}=\prod_{i} \otimes \Psi_{\alpha_{i}} \mbox{ with } \sigma^{x}_{i}\Psi_{\alpha_{i}}=\alpha_{i}\Psi_{\alpha_{i}}
\eqno{(15)}
$$
with $\alpha_{i}=\pm1$, in which $\rho(0)$ is diagonal, we see that the terms in (14) which contribute to the right hand side of (6) are (using $\sigma^{z}\sigma^{x}=i\sigma^{y}$ et cicl.):
$$
((\cos(\epsilon(j)t))^{2}-(\sin(\epsilon(j)t))^{2})\rho(0)(\sigma^{x}_{i_0})\cos(2\epsilon(j)t)=\\
=(\cos(2\epsilon(j)t))^{2}\phi_{i_0}(\sigma^{x}_{i_0})
\eqno{(16)}
$$

The above formula is true, however, only if $i_0-j\ge -n$, $i_0+j\le n$ in (14), due to the restriction in (10). If, e.g.,
 $i_0-j<-n$, with $i_0+j<n$, the first and third terms in (14) will be replaced by the identity, and the resulting term,
instead of (16), will be
$$
\cos(2\epsilon(j)t)\phi_{i_0}(\sigma^{x}_{i_0})
\eqno{(17)}
$$

For this reason it is important to perform the limit on the r.h.s. of (7) before the one on the r.h.s. of (8): the asymmetry  which yields a few terms of the form (17) in the case of a finite system for $i_0\ne0$ disappears upon taking the limit 
 $n\to\infty$. For this reason it is also important to take the infinite volume limit in (7) in the sense of Fisher, which means roughly that the domains $V$ have to increase at about the same rate in all directions (see [8] for comments and references): in one dimension, this condition is satisfied by the symmetric choice in (10).

With the above remarks, we are allowed to restrict ourselves to $i_0=0$ in (7) and obtain, with (10), (3):
$$
\langle\sigma^{x}_{0}\rangle_{n}(t)=\phi_{0}(\sigma^{x}_{0})\prod_{j=1}^{n}(\cos(2\epsilon(j)t))^{2}\times\cos(2Bt)
\eqno{(18)}
$$
We now have:

\textbf{Proposition2.1}
Let
$$
\phi_{0}(\sigma^{x}_{0})\ne0 
\eqno{(19)}
$$
Then, if $\epsilon\in l_2$, i.e., (12b) holds,
$$
\exists\lim_{n\to\infty}\langle\sigma^{x}_{0}\rangle_{n}(t)\equiv\langle\sigma^{x}_{0}\rangle(t)
\eqno{(20)}
$$
If $\epsilon\in l_1$, i.e., if (12a) holds, for no constants $0<\alpha<\infty$,$0<C<\infty$, the inequality
$$
\vert \langle\sigma^{x}_{0}\rangle(t) \vert\le C \exp(-\vert \alpha \vert t) \mbox{ if } t\ge 0
\eqno{(21)}
$$
is satisfied.

\textbf{Proof}: By (12b) and (11b), the product on the r.h.s. of (18) converges to the infinite product
$$
\wp\equiv \prod_{j=1}^{\infty}(\cos(2\epsilon(j)t)^{2})
\eqno{(22a)}
$$

Above, $\wp$ exists by the following simple bound: by (11b), for any $t$ there exists $M<\infty$ such that 
$2 \vert \epsilon(j) \vert t <1\forall j\ge M$ and thus
$$
\wp\le d\prod_{j=M}^{\infty}(\vert \cos(2t(\epsilon(j))^2) \vert)\le d\prod_{j\ge M}(1-ct^{2}(\epsilon(j))^2)\\
\le d\exp(-ct^2\sum_{j\ge M}(\epsilon(j))^2)
\eqno{(23)}
$$

Above, $c$ and $d$ denote positive constants which may vary at each step and we used the elementary inequalities:
$ 0< \cos(x) <1-cx^2 $, $ 1-x< \exp(-x) $, for some $c>0$ if $0<x<1$. We have thus proved (20), with

$$
\langle\sigma^{x}_{0}\rangle(t)= \phi_{0}(\sigma^{x}_{0})\wp
\eqno{(22b)}
$$
By ([4],pg.2954, and Proposition 1):
$$
\langle\sigma^{x}_{0}\rangle(t)=(\Pi_{\rho}(\sigma^{x}_{0})\Phi_{\rho},\exp(it\Pi_{\rho}(2\tilde{H_0})\Phi_{\rho})
\eqno{(24)}
$$
where
$$
\tilde{H_0}=\sigma^{z}_{0}\sum_{k=-\infty}^{\infty} \epsilon(\vert k \vert)\sigma^{z}_{k}
\eqno{(25)}
$$

Above, $\tilde{H_0}$ is the norm limit, as $n\to\infty$ (in the quasi-local algebra $\cal{A}$ [11]) of the sequence of elements $\sigma^{z}_{0}\sum_{k=-n}^{n}\epsilon(\vert k \vert)\sigma^{z}_{k}$, which is the generator of the group of time-translation automorphisms of $\cal{A}$, and $\rho=\rho_{\infty}$ the state on $\cal{A}$, $\rho=\otimes_{j\in\mathbf{Z}}\phi_{j}$ corresponding to (3a), with $\Pi_{\rho}$ the GNS representation associated to $\rho$ with cyclic vector $\Phi_{\rho}$ [11].
By (25), $\tilde{H_0}$ is bounded below (in fact it is bounded) and (21) follows from a standard result in function theory (see,e.g., [10], § 5.8).

\textbf{Remark 2.1}

The proof of non-exponential decay relies on the form (25), which is meaningless when only (12b) (but not (12a)) is valid, and, indeed, we see from (23), by choosing $M\ge t^{1/\alpha}$, that, for potentials of class $l_2$ of the form (48a),with $1/2<\alpha\le1$, exponential decay (in the sense of inequality (21)) is possible when only (12b) is required. The latter suffices to prove (20), as we saw, but it is clear from (2) or (10) that it precludes thermodynamic stability for the present model. We shall see in the next section that this is remedied by the introduction of randomness.

\section{ The random model}

We now introduce the Hamiltonian of a disordered system corresponding to (10) (with $B=0$ for simplicity):
$$
\tilde{H_{n}}=1/2\sum_{j,k=-n}^{n} J(j,k)\epsilon(\vert j-k \vert)\sigma^{z}_{j}\sigma^{z}_{k}
\eqno{(28)}
$$
where $J(j,k)$ are, again for simplicity, independent, identically distributed random variables (i.i.d. r.v.). We shall use  $Av(\cdot)$ to denote averaging with respect to the random configuration ${J}$. The ${J}$ are assumed to satisfy [8]:
$$
Av(J(j,k))=0
\eqno{(29a)}
$$

$$
\vert Av(J^{n}(j,k)) \vert \le n!c^{n} \forall n=2,3,4,\cdots
\eqno{(29b)}
$$

Let the free energy per site $f_{n}$ be defined by
$$
f_{n}(J)\equiv \frac{-kT}{2n+1}\log Z_{n}(J)
\eqno{(30a)}
$$
where
$$
Z_{n}(J)= tr(\exp(-\beta\tilde{H_{n}}))
\eqno{(30b)}
$$
is the partition function and the trace is over the Hilbert space (2). Then

\textbf{Theorem 3.1} [9]
Under assumptions (12b) and (29), the thermodynamic limit of the free energy per site
$$
f(J)=\lim_{n\to\infty} f_{n}(J)
\eqno{(31a)}
$$
 exists and equals its average:
$$
f(J)= Av(f(J))=\lim_{n\to\infty} Av(f_{n}(J))
\eqno{(31b)}
$$
for almost all configurations $J$ ($\mbox{ a.e. } J $).

In [9] general $\nu$ dimensional systems were considered (as remarked, we could equally have done so here), with the limit of finite regions taken in Fisher's sense. The reason why (12b) suffices for the existence of the thermodynamic limit is that, in order to obtain a uniform lower bound for the average free energy per site, the cumulant expansion (see, e.g., ([13], (12.14), pg 129, for the definition of $Av_c$, there called Ursell functions):
$Av(\exp(tJ(i,j))=\exp(\sum_{n=2}^{\infty}Av_c(J^{n}(i,j))t^{n}/n!)$ was used [8], which, by (29a), starts with the second cumulant $Av_c(J^{2}(i,j))=Av(J^{2}(i,j))$, which is the variance of $J(i,j)$. Condition (29b) was used to control the sum in the exponent above. 
 
We now investigate whether model (28), which is thermodynamically stable in the sense that the thermodynamic limit for the free energy exists under assumption (12b) for the lattice potential by Theorem 3.1, also yields a well defined thermodynamic limit for the dynamics, i.e., of $\langle\sigma^{x}_{i_0}\rangle(t)$. Instead of (14), a typical term containing the interaction $\epsilon(j)$ in the operator $U^{-t}_{n}\sigma^{x}_{i_0} U^{t}_{n}$ appearing on the r.h.s. of (6) is, now,
\begin{eqnarray*}
(\cos(tJ_{i_0-j,i_0}\epsilon(j))-i\sigma^{z}_{i_0}\sigma^{z}_{i_0-j}\sin(J_{i_0-j,i}\epsilon(j)t))& \times &
(\sigma^{x}_{i_0}\cos(2J_{i_0,i_0+j}\epsilon(j)t)+\sigma^{y}_{i_0}\sigma^{z}_{i_0+1}\sin(2J_{i_0,i_0+j}\epsilon(j)t))\\
& \times &(\cos(J_{i_0-j,i_0}\epsilon(j)t)+i\sigma^{z}_{i_0}\sigma^{z}_{i_0-j}\sin(J_{i_0-j,i_0}\epsilon(j)t))
\end{eqnarray*}
$$\eqno{(32)}$$

We ignore the differences commented after (15), which are analogous in the present case but, again, will disappear upon taking the limit $\lim_{n\to\infty}$ as discussed there. For any fixed configuration ${J}$, an estimate of the same type (23) in proposition 2.1 yields a formula analogous to (22b) in the limit $n\to\infty$:
$$
\langle\sigma^{x}_{i}\rangle(t)=\rho^{t}_{\infty}(\sigma^{x}_{i})= \phi_{i}(\sigma^{x}_{i})\wp_{i}
\eqno{(33)}
$$
where
$$
\wp_{i}\equiv\prod_{k=1}^{\infty}\cos(2tJ_{i,i+k}\epsilon(k))\cos(2tJ_{i-k,i}\epsilon(k))
\eqno{(34)}
$$
and we use the same notation as in (8). We now obtain for the finite version of the r.h.s. of (8), with $\Lambda=[-m,m]$
 and $N(\Lambda)=2m+1$:

$$
\rho^{t}_{\infty}(m)\equiv\rho^{t}_{\infty}(\frac{\sum_{i=-m}^{m}\sigma^{x}_{i}}{2m+1})=\\
=\delta(\frac{\sum_{i=-m}^{m}\wp_{i}}{2m+1})
\eqno{(35)}
$$
where
$$
\delta=\phi_{i}(\sigma^{x}_{i})
\eqno{(36)}
$$
is a nonzero number independent of $i$. We have now:

\textbf{Theorem 3.2}
Let the r.v. ${J}$ be Bernoulli or Gaussian. Then there exists a subsequence $m_{r=1,\infty}$ such that
$$
\lim_{r\to\infty}\rho^{t}_{\infty}(m_{r})= \tilde{\wp_{i_0}}\equiv Av(\wp_{i_0})
\eqno{(37)}
$$
$\mbox{ a.e. }J$.

\textbf{Proof.}
Consider the $(2m+1)\times (2m+1)$ matrix with the r.v. $\wp_{i}$ as entries. $\wp_{i}$ and $\wp_{j}$, for $i\ne j$, are not independent. For, let $j>i$. Then,
$$
\wp_{i}\equiv \prod_{k=1}^{\infty} \cos(2tJ_{i,i+k}\epsilon(k))\cos(2tJ_{i-k,i}\epsilon(k))
\eqno{(38a)}
$$
and
$$
\wp_{j}\equiv \prod_{k=1}^{\infty} \cos(2tJ_{j,j+k}\epsilon(k))\cos(2tJ_{j-k,j}\epsilon(k))
\eqno{(38b)}
$$
If $i+k=j$, $j-k=i$, and there will be one common factor 
$\cos(2tJ_{i,j}\epsilon(k))$ in $\wp_{i}$ and $\wp_{j}$. We now consider the r.v. on the r.h.s. of (35):

$$
X_{m}\equiv\frac{\sum_{i=-m}^{m}\wp_{i}}{2m+1}
\eqno{(39)}
$$
Since the $J_{i,j}$ are identically distributed, $\tilde{\wp_{i}}$ is independent of $i$. Consider
$$
\tilde{X_{m}}\equiv\frac{\sum_{i=-m}^{m}(\wp_{i}-\tilde{\wp_{i}})}{2m+1}
\eqno{(40)}
$$
We have:
$$
\lim_{\vert i-j \vert \to\infty}Av((\wp_{i}-\tilde{\wp_{i}})(\wp_{j}-\tilde{\wp_{j}}))=0
\eqno{(41)}
$$
To prove (41), let $j>i$ for definiteness. By (38), with $k=j-i$,
$$
\vert Av((\wp_{j}-\tilde{\wp_{j}})(\wp_{k}-\tilde{\wp_{k}})) \vert\le\\
\mbox{ const. } \vert Av(\cos(2J_{i,i+k}\epsilon(k)t))^2-(Av(\cos(2J_{i,i+k}\epsilon(k)t)))^2 \vert
\eqno{(42)}
$$
The r.h.s. of (42) is identically zero for Bernouilli r.v.. For the Gaussian,
$$
Av((\cos(2J_{i,i+k}\epsilon(k)t)^2)= \frac{\int_{-\infty}^{\infty}\exp(-x^2)(\cos(2x\epsilon(k)t))^2}{\sqrt(\pi)}\\
=\frac{1+4\exp(-t^2(\epsilon(k))^2)}{2}
\eqno{(43a)}
$$
while

$$
Av((\cos(2J_{i,i+k}\epsilon(k)t))^2)=\exp(-2(\epsilon(k))^2 t^2)
\eqno{(43b)}
$$
Putting (43) into (42) we find that the r.h.s. of (42) is $O((\epsilon(k))^4)$ for large $k$, which tends to zero by (11b) and (12b), thus proving (41). By (40) and (41),
$$
\lim_{m\to\infty}Av((\tilde{X_{m}})^2)=0
\eqno{(44)}
$$
Thus, by Chebyshev's inequality, $\tilde{X_{m}}\rightarrow 0$ in probability, which implies by the Borel-Cantelli lemma that a subsequence converges to zero $\mbox{ a.e. }{J}$ (see, e.g., [12], Theorem 4.2.3). This proves (37).

The convergence $\mbox{ a.e. } {J}$ is the well-known property of self-averaging, which we conjecture holds even without the necessity of restricting to a subsequence. As discussed in [8] and more extensively in [13] (``Thou shalt not average'') it is essential for the reproducibility of the outcomes of experiments, in this case the measurement of the transverse magnetization.

By (8), (37) and (43b), we have the results:
$$
f_B(t)=\prod_{k=1}^{\infty}(\cos(2t\epsilon(k)))^2
\eqno{(45a)}
$$
for the Bernoulli distribution and 
$$
f_G(t)=\exp(-2t^2 \sum_{k=1}^{\infty}(\epsilon(k))^2)
\eqno{(45b)}
$$
for Gaussian r.v..

\textbf{Remark3.1}
By (45a) and (22a,b), the rate of approach to equilibrium agrees with the nonrandom rate in the Bernoulli case. This leads immediately to the interpretation of (22a,b) for lattice potentials of class $l_2$, i.e., satisfying (12b), in view of Theorem 3.1. See also remark 2.1.

\section{The rates of approach to equilibrium}

By (45b) $f_G$ is explicit in the Gaussian case:
$$
f_G(t)=\exp(-2t^2 \sum_{k=1}^{\infty}(\epsilon(k))^2)
\eqno{(46)}
$$
From (46) it is exponential in the sense of inequality (21) whatever the range of the interaction.
We have, now:

\textbf{Proposition 4.1}
The approach to equilibrium for Gaussian r.v. is faster than for the Bernoulli distribution in the case of potentials of class $l_2$, i.e.,:
$$
\lim_{t\to\infty}\frac{f_B(t)}{f_G(t)}=\infty
\eqno{(47)}
$$
if
$$
\epsilon(k)=k^{-\alpha} \mbox{ with } 1/2<\alpha\le1 \mbox{ and } k\ge1
\eqno{(48a)}
$$
The same holds in the prototypical example of infinite, but extreme short range case
$$
\epsilon(k)=2^{-k-1} \mbox{ with } k\ge1
\eqno{(48b)}
$$

\textbf{Proof}
We first consider (48a) and use the elementary inequalities:
$$
\cos(x) >  1 - x^2/2  
\eqno{(49a)}
$$
valid for $0<x<1$ and
$$
\exp(-x)  <  1- x/2  
\eqno{(49b)}
$$
valid for $0<x<1.5936$, to find a lower bound to $f_B(t)$, given by (45a). We assume that the time values are sampled at points differing from the zeros of the cosine function in both cases (48a) and (48b). Then, by (49),

\begin{eqnarray*}
\vert f_B(t) \vert = \vert \prod_{k=1}^{\infty}(\cos(2tk^{-\alpha})^2 \vert\ge\\
\mbox{ const. }(2t)^{1/\alpha}\prod_{2tk^{-\alpha}<1}(1-1/2(2tk^{-\alpha})^2)^2 & \ge &
\mbox{ const. }(2t)^{1/\alpha}\exp(\sum_{2tk^{-\alpha}<1}8t^2k^{-2\alpha})\\
& \ge &\mbox{ const. }(2t)^{1/\alpha}\exp(-8t^2\int_{(2t)^{1/\alpha}}^{\infty}\,dxx^{-2\alpha})\\
& = &\mbox{ const. } \exp(-d t^{1/\alpha}+ c \log(t))
\end{eqnarray*}$$\eqno{(50)}$$
where $d,c$ are positive constants. (47) follows from (46), (48a) and (50).
We now turn to (48b), for which an exact result exists for $f_B(t)$:
$$
f_B(t)=\prod_{k=1}^{\infty}(\cos(2t/2^{k+1}))^2=\prod_{k=1}^{\infty}(\cos(t/2^{k}))^2=(\sin(t)/t)^2
\eqno{(51)}
$$
by Vieta's formula

$$
\frac{\sin(t)}{t}=\prod_{k=1}^{\infty}\cos(t/2^{k})
\eqno{(52)}
$$
and (47) is obvious from (46) and (51).

As a last remark concluding this section, Vieta's remarkable formula (52) may be viewed from the point of view of characteristic functions and hence of addition of independent r.v. in probability theory: note that the l.h.s. of (51) is the characteristic function of the uniform distribution in $[-1,1]$ ([12], exercise 8, pg.165, or [14], Chap. 1). It admits, however, an elementary proof, discovered by S.P.Heims([28], Appendix): using the identity
$\sin\alpha=2\sin(\alpha/2)\cos(\alpha/2)$ successively, we obtain 
$\sin\xi=2^{n}\sin(\xi/2^{n})\cos(\xi/2^{n}\cos(\xi/2^{n-1})\cdots\cos(\xi/2)$. But 
$\lim_{n\to\infty}\frac{\sin(\xi/2^{n})}{\xi/2^{n}}=1$, from which (52) results.

\section{Conclusion and open problems}

We have shown that the inclusion of randomness permits a natural generalization of the Emch-Radin model to the class $l_2$ of square-summable lattice potentials. The generalized model (as well as the original one, of course) provides one of the very few examples of a random quantum system allowing a rigorous analysis of the dynamics of approach to equilibrium, and, in particular, of the role probability distributions play there.

In [15], together with S.R. Salinas, we studied the phase diagram of the mean-field random Ising model, and proved that a tricritical point exists there for the Gaussian but not in case of a Bernoulli distribution. Thus, equilibrium properties are, in general, also quite sensitive to the nature of the probability distribution. We suggested [15] that the observed differences might be due to the fact that a discrete distribution of probabilities samples just a few values of the couplings and therefore introduces some short-ranged elements into the problem (see also [16]). From this point of view, it may be conjectured that discrete distributions have a closer connection with real materials and, indeed, in the nonrandom case - which is a caricature of a real old experiment [27] - a decay of type (52) provides a better qualitative description than (48).
The issue can only be settled by performing analogous experiments on random systems, as briefly discussed below. In particular, recent great progress in spin-glass theory (see [17] and references given there) pertains to the mean-field SK theory, which is long-range. For these, gaussian distributions may be more adequate, since they sample many values of the couplings, and ``reinforce'' in this sense the long-range nature of the interactions. For such systems, proposition 4.1 predicts a faster approach to equilibrium.

Concerning the relation to experiment, the experiments at the University of Chicago and ATT Bell Laboratories [29] have shown the importance of the quantum regime in the study of the phase diagram of certain materials. In the choice of [29], a strong spin-orbit coupling between the spins of the magnetic ion component of the material and the underlying crystal essentially restricts the spins to orient either parallel or antiparallel to a specific crystalline axis: such spins are usually referred to as Ising spins. Under application of a transverse magnetic field, oriented perpendicular to the preferred axis, a flipping effect of the Ising spins takes place, analogous to the one observed in a spin-resonance experiment, and it becomes possible to identify a quantum crossover regime at sufficiently low temperatures and certain (critical) values of a parameter proportional to the transverse field intensity. This has been nicely reviewed by S. Sachdev in [30]. The qualitative similarity to the present model is obvious, but doing an experiment on such materials under conditions analogous to the old experiment reported in [28] seems to be a very interesting open problem.

We finally come to the interesting conceptual question of exponential versus nonexponential decay in the sense of inequality (21). By (48), exponential decay occurs in the Gaussian case for interactions of both short and long range, and in the Bernoulli case for very long-range interactions of class $l_2$ by (23). On the contrary, for nonrandom systems exponential decay is excluded by proposition 2.1, in contrast to the behavior of the recurrence time probability in the classical mechanics of fully chaotic systems (see, e.g., [18], for a review and references).

For one particle in a short-range potential in quantum mechanics, algebraic decay $O(t^{-\nu/2})$, where $\nu$ is the dimension, is the rule, as known from scattering theory, due to the spreading of the free wave-packets (see, e.g., [19]). In contrast, algebraic decay results from mixed phase space and trapping of trajectories in regular islands [18]. However, for quantum systems in external (non-decaying) fields, such as constant electric fields [20] or parametric oscillators [21], exponential decay does occur: the Hamiltonian is not bounded below, i.e., its spectrum is the whole line and proposition 2.1 does not go through. Such models, dubbed quantum Anosov systems in [22], are characterized by positive quantum Lyapunov exponents ([20],[21],[22]). In the present model, the existence of exponential behavior is due to the randomness, while in the cited systems the non-semiboundedness of the Hamiltonian arises from neglecting the back-reaction of the field.

A final point is that, in spite of the simplicity of the present model, it is a genuine quantum many-body problem, i.e., of infinite number of degrees of freedom, about which almost no results on time evolution with good ergodic properties exist: one exception is [23]. There it is pointed out that the condition for existence of a state-dependent time-evolution in the Heisenberg picture is thermodynamic stability in the sense of the N-particle Hamiltonian having a lower bound of the form 
$- \mbox{ const. }N$, which would require, in our case, lattice potentials of class $l_1$.. Accordingly, the Heisenberg time evolution is well-known to exist as an automorphism of the quasi-local algebra in the $l_1$ case (see [4] and [11]), i.e., in a state-independent sense, and it is at best state-dependent in the $l_2$ case: an example is the present construction, which is valid for the same class of states considered in ([3],[4]).

Acknowledgement: We thank the referees for relevant remarks and corrections.

\end{document}